\documentstyle[aps,twocolumn,prl,tighten,flushrt]{revtex} 
\input epsf
\def\plotone#1{\centering \leavevmode
\epsfxsize= 0.7\columnwidth \epsfbox{#1}}

\def\be{\begin{equation}}
\def\ee{\end{equation}}
\def\bea{\begin{eqnarray}}
\def\eea{\end{eqnarray}}

\def\cmm2{{\,\rm cm^{-2}}}
\def\cm2{{\,{\rm cm}^2}}
\def\cmm3{{\,{\rm cm}^{-3}}}
\def\gcmm3{{\,{\rm g\,cm^{-3}}}}

\def\fun#1#2{\lower3.6pt\vbox{\baselineskip0pt\lineskip.9pt
  \ialign{$\mathsurround=0pt#1\hfil##\hfil$\crcr#2\crcr\sim\crcr}}}

\def\vec{\bf}

\def\msun{{\,M_\odot}}

\def\'{^{\prime}}

\def\p3m{P$^3$M}

\def\ga{\mathrel{\mathpalette\fun >}}
\def\fun#1#2{\lower3.6pt\vbox{\baselineskip0pt\lineskip.9pt
  \ialign{$\mathsurround=0pt#1\hfil##\hfil$\crcr#2\crcr\sim\crcr}}}

\font\BF=cmmib10
\def\gam{\hat{\gamma}}

\def\v{{\hbox{\BF v}}}
\def\d{\delta}

\begin{document}
\twocolumn[\hsize\textwidth\columnwidth\hsize\csname @twocolumnfalse\endcsname
\preprint{CfPA/97-th-11,CITA-97-31}
\title{The Impact of Inhomogeneous Reionization on\\ 
Cosmic Microwave Background Anisotropy}
\author{Lloyd\ Knox$^1$, Rom\'an\ Scoccimarro$^1$, and Scott\ Dodelson$^2$}
\address{$^1$ Canadian Institute for Theoretical Astrophysics, Toronto, ON M5S 3H8, CANADA}
\address{$^2$ NASA/Fermilab Astrophysics Center\\
Fermi National Accelerator Laboratory, Batavia, IL 60510, USA}
\date{\today}
\maketitle

\begin{abstract}
It is likely that the reionization of the Universe 
did not occur homogeneously.
Using a model that associates
ionized patches with overdense regions, we find that the resulting
cosmic microwave background (CMB) anisotropy power spectrum peaks at
angular scales corresponding to the extent of the ionized regions, and
has a width that reflects the correlations between them. There is
considerable uncertainty in the amplitude. Neglecting inhomogeneous
reionization in the determination of cosmological parameters from high
resolution CMB maps may cause significant systematic error.
\end{abstract}
\pacs{Valid PACS appear here.}
]
{\parindent0pt\it Introduction.} 
Observations of cosmic microwave background (CMB) anisotropy
are providing strong constraints on theories of cosmological
structure formation.  Planned CMB observations
can potentially provide constraints on the parameters of these theories 
at the percent level\cite{forecast,BET}.  
One of the reasons the CMB is such a wonderful probe of cosmological
parameters is that predictions for a given theory can be made with
great precision\cite{cmbfast}.  This is because linear perturbation
theory is an excellent approximation over the relevant length scales
at $z \simeq 1100$ when the CMB was last tightly coupled to matter.

However, new anisotropies can be generated at late times 
by the non-linear process of reionization as the 
CMB photons are brought back into
contact with matter via Thomson scattering.  We know this reionization
took place before redshift $z\simeq 5$ because spectra of distant
quasars do not show a continuum of absorption by neutral hydrogen.  It is
unlikely that reionization occurred at $z \ga 60$ for in that case the
level of anisotropies on degree scales would be significantly lower
than on ten degree scales, when actually
just the opposite is true \cite{peakexists}.

Reionization affects the CMB in several ways.  First,  
the fluctuation power on small angular scales is damped by a
factor $e^{-2\tau}$ where $\tau$ is the optical depth back to last
scattering. Second, as Sunyaev and Kaiser \cite{Kaiser} pointed out,
secondary anisotropies are generated on large scales due to
the Doppler effect when photons scatter off moving free electrons.
They also noted that the effect is strongly suppressed on small scales
because photons get nearly opposite Doppler shifts on different sides of a
density peak, a consequence of potential flows generated by
gravitational instability. Both of these effects---the damping and the
Sunyaev-Kaiser (SK) effect---are linear and therefore included in
standard Boltzmann codes\cite{cmbfast}.

Here we focus
on the effects of inhomogeneities in the ionization 
field \cite{ADPG,GH,pj98}.  
We show that inhomogeneous reionization (IHR) 
restores the SK effect at small scales due to modulation of the
velocity field with the spatial variation of the ionization
fraction,  an effect analogous to the modulation by 
the density field  
in the homogeneous case at second order in perturbation theory\cite{Vishniac}.  Since 
the typical patch size is much smaller than the scale
of variation of the velocity field, the Doppler effect
contributions come only from regions where the flow is coherent.
The resulting anisotropy pattern
depends sensitively on the model of reionization. 
Thus, the signature of IHR in the CMB provides an indirect
window on a poorly probed but interesting era in which the first
stars and galaxies formed.  On the other hand, if these 
anisotropies are large enough, they can create systematic
errors in the determination of cosmological parameters. 
So, after calculating the anisotropies, we compute the magnitude
of these systematic errors, i.e. how much a given
cosmological parameter might be {\it misestimated} if IHR is ignored.

{\parindent0pt\it Anisotropy Spectrum: Formal Solution.}
The perturbation to the photon temperature distribution function,
$\Delta \equiv \delta T/T$, is governed by the Boltzmann equation, which
at the late times of interest is 
\be
\label{eq:boltz}
\partial \Delta({\bf x},\eta,\gam)/ \partial \eta = n_e \sigma_T a 
\gam\cdot \vec v(\vec x,\eta)
\ee
where $a=(1+z)^{-1}$ is the scale factor of the universe normalized to 
unity today,
$\eta \equiv
\int dt/a$ is the conformal time, $n_e$ is the free electron 
density, $\sigma_T$ is the Thomson cross-section, $\gam$
is the direction of the photon momentum, and $\vec v$ is the electron 
velocity. Henceforth we work in units
where the conformal time today is unity, $\eta_0=1$; and assume a
flat, matter dominated universe, $a = \eta^2$.
The solution to Eq.~\ref{eq:boltz} for the photon
perturbations here and now 
(at ${\vec x} = {\vec x}_0$ and $\eta= 1$;
$\Delta_0(\gam ) \equiv \Delta({\bf x}_0,\eta_0,\gam )$\ ) 
is
\be
\label{eq:soln}
\Delta_0(\gam ) 
=  4.06\times 10^{-5}\ \Omega_b h  
\int_{\eta_i}^{1} \frac{d\eta}{\eta^3}\  
x_e({\bf x},\eta)\  \gam\cdot \v ({\bf x}),
\ee
where, ${\vec v}({\bf x},\eta) \equiv -2\eta\ \v({\bf x})$ (Mpc/h),
$n_e=\bar n_e\ x_e({\bf x},\eta)a^{-3}$, and   
$\bar n_e \sigma_{\rm T} \eta_0 = 0.122 \Omega_b
h$, with $\bar n_e$ the present mean electron density, $\Omega_{b}$ 
the baryon density in units of the critical density and
$h$ the Hubble constant in units of 
100 km s$^{-1}$ Mpc$^{-1}$.  Both $x_e$ and $\v$ are 
evaluated at ${\bf x} = {\bf x}_0 - \gam(1 - \eta)$, 
the position at time $\eta$ of a photon with direction
vector $\gam$ incident on us today. 
Note that the integral includes all contributions starting from $\eta_i$,
a time far after standard recombination at $\eta \simeq 0.03$, but before
reionization occurs.

We focus on predicting the two-point correlation 
function, $C(\theta) \equiv \langle
\Delta_0(\gam) \Delta_0(\gam') \rangle$ where 
$\cos\theta \equiv \gam\cdot\gam'$ and the angular brackets 
indicate an average over all locations
${\bf x}_0$.
Using Eq.~\ref{eq:soln} we obtain 
 the correlation function due to reionization:  
\be
\label{eq:cihr}
C(\theta) =
10^{-12} \Big(\frac{\Omega_b h}{0.025}\Big)^2 \int_{\eta_i}^1
\frac{d\eta}{\eta^3}  
\int_{\eta_i}^1 \frac{d\eta'}{\eta'^3} {\cal C}(\eta,\eta',\theta).
\ee
It is therefore
determined by an integral over adjacent lines of sight of the
four-point correlation function
\be
\label{eq:defc}
{\cal C}(\eta,\eta',\theta) \equiv \gam_i\gam'_j
\ \langle x_ex_e^\prime v_i v'_j \rangle,
\ee
where, for example, $x_e^\prime \equiv x_e(\vec x',\eta')$. 
Different models of reionization will lead to different
four-point functions. We now explore several possibilities.

{\parindent0pt\it Homogeneous Reionization.}
The simplest possibility is that reionization takes
places homogeneously. While not particularly plausible, it
is a useful limiting case for demonstrating the
SK cancellation. In homogeneous
reionization (HR), $x_e \equiv \bar x_e(\eta)$, where $\bar x_e$ is
fraction of the Universe that is ionized. Thus we have:
\be
{\cal C}^{\rm HR}(\eta,\eta',\theta)   
=  \bar x_e \bar x_e^\prime \ C_v(\eta,\eta',\theta)
\ee
and we take $\bar x_e$ to be zero at times earlier than
$z_i+\delta z_i$, 
unity at times later than $z_i$ and to increase linearly
with redshift during the transition.
The velocity correlation function can be written as
\be
\label{eq:cv}
C_v(\eta,\eta',\theta) =
\Psi_{\perp} \cos\theta +
\left( \Psi_{\parallel} - \Psi_{\perp} \right)
{({\vec r} \cdot \gam)\ ({\vec r} \cdot
\gam') \over r^2}, 
\ee
where $\vec r\equiv \vec x- \vec x'$. The correlation functions of the
velocity components parallel and perpendicular to $\vec r$ are
 \bea
 \Psi_{\parallel}(r) &=& \int \frac{P(k)}{k^{2}}\ \Big[j_{0}(kr)-
 2 \frac{j_{1}(kr)}{kr}\Big]\ d^{3}k, \\
 \Psi_{\perp}(r) &=& \int \frac{P(k)}{k^{2}}\ \frac{j_{1}(kr)}{kr}\
 d^{3}k,\eea
where $P(k)$ is the matter power spectrum, throughout assumed
to be that of the standard cold dark matter theory \cite{BBKS}. 
Fig. 1 shows $C_v(\eta,\eta',0)$ as a function of $\eta$ for fixed
$\eta'=0.188$. $C_v$ is strongly peaked at equal times, where the two
lines of sight share the same velocity. On the other hand, at large
$|\eta-\eta'|$, $C_v$ becomes negative due to infall from opposite 
sides into an overdense region. 
The integration, in Eq.~\ref{eq:cihr}, of this oscillatory function with 
the slowly-varying $\bar x_e \bar x_e^\prime$
leads to a cancellation of Doppler effects, as described by 
Kaiser\cite{Kaiser} in Fourier space. 

{\parindent0pt\it Uncorrelated Inhomogeneous Reionization.}  The SK
cancellation can be avoided if the velocity two-point function $C_v$
is modulated by the spatial dependence of $x_e$.
To demonstrate this effect, we adopt a toy model
of the reionization process:  
independent sources turn on randomly and instantaneously
ionize a sphere with comoving radius $R$ and volume $V_R$, which
then remain ionized.  The resulting four-point function can be
written as:
\be
\label{eq:uc1}
{\cal C}^{\rm uc}(\eta,\eta',\theta) =C_v(\eta,\eta',\theta)\  
\left[C_{x_e}^{\rm uc}(\eta,\eta',\theta)+\bar x_e \bar x_e^\prime\right].
\ee
Since $\bar \chi$ (where $\chi \equiv 1-x_e$) is equal to the
probability that no ionizing source is within volume $V_R$,
the probability that no source is within $R$ of either
${\bf x}$ or ${\bf x'}$ is (at equal times) $\langle \chi \chi' \rangle =
\bar \chi^{(2V_R-V(r))/V_R} = \bar \chi \bar \chi \bar \chi^{-V(r)/V_R}$
where for $r<2R$, $V(r)=V_R\ [1 - (3/4)\ (r/R) + (1/16)\ (r/R)^3]$ is 
the overlap volume of the spheres centered on ${\bf x}$ and ${\bf x'}$.
Therefore, with the proper generalization to unequal times, we have:
\bea
C_{x_e}^{\rm uc} & \equiv & \langle x_e x_e^\prime \rangle - \bar x_e \bar x_e^\prime =
\langle \chi \chi^\prime \rangle - \bar \chi \bar \chi' \\
 & = & \bar \chi \bar \chi'\left[\bar 
\chi({\rm min}(\eta,\eta'))^{-V(r)/V_R} - 1\right]
\label{eq:uc2}
\eea

\begin{figure}[bthp]
\plotone{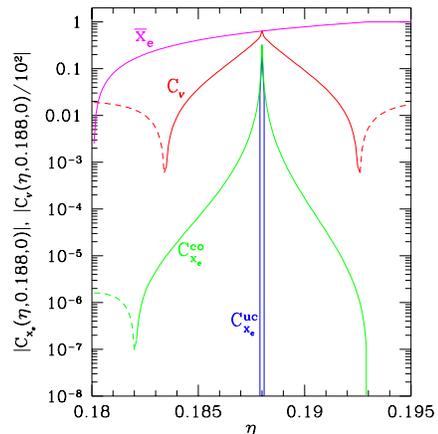}
\caption[caption]{The mean ionization fraction $\bar x_e$ for HR
($z_i=26$, $\d z_i=4$), the  velocity two-point function
$C_v(\eta,0.188,0)$ for the SCDM (standard cold dark matter) 
model with $\sigma_8=1.2$ where $\sigma_8$ is the rms of mass fluctuations
in spheres of radius $8h^{-1} $\ Mpc today, and the
ionization fraction two-point function for 
uncorrelated $C_{x_e}^{\rm uc}$ and correlated bubbles
$C_{x_e}^{\rm co}$. Dashed lines denote negative values.} 
\label{fig1}
\end{figure}

The resulting two-point function, $C_{x_e}^{\rm uc}$, is sharply peaked
at small $r\approx|\eta-\eta'|$ (see Fig. 1, where $R=1$ Mpc), 
therefore in Eq.~\ref{eq:uc1}, ${\cal C}^{\rm uc}(\theta) \propto  
C_v(\eta,\eta,\theta)$.   
Thus, the SK cancellation of homogeneous reionization is avoided;
IHR modulates 
the velocity field, accessing only the region where the velocity field
is highly coherent. Note though that the modulation is not particularly
effective since the ionization radius $R$ is typically very small, and
thus only a tiny region of $\eta$ contributes. Therefore, as 
Gruzinov and Hu (GH)\cite{GH} have pointed out, a model
of this type necessarily produces anisotropies which have 
amplitude proportional to $R$. 

Figure 2 shows the anisotropy spectrum arising from the two-point
function of Eq.~\ref{eq:uc2}, assuming 
$z_i=26$ and $\d z_i=4$.
The GH results for the same parameter choices  
would each
have peaks 30\% lower in $l$ and two times larger in amplitude due
to a slower fall off of $C_{x_e}^{\rm uc}$ with $r$ and 
one less factor of $(1-\bar x_e)$ in their version of Eq.~\ref{eq:uc2}.
Despite these differences, 
they correctly identified the qualitative features of this type of 
IHR: (i) amplitude proportional to $R$; (ii)
white noise ($C_l^{\rm IHR} \propto $ constant) on large scales due to
lack of correlations at $r \gg R$; and (iii)
peak at $l \sim (1 - \eta_{r}) / R$ where reionization takes place
at $\eta_r$. We now show that the first two of these
features break down when more realistic models are considered.

\begin{figure}[bthp]
\plotone{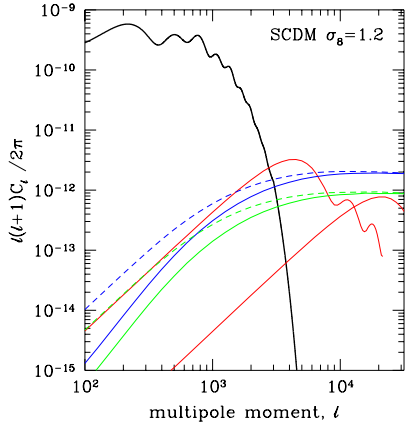}
\caption[caption]{The angular power spectrum, 
$C_l$, defined by $C(\theta) \equiv
\sum_l (2l+1) C_l/(4\pi)P_l(\cos \theta)$. The thick solid line 
on top is the linear spectrum.
The rest are different approximations to the IHR anisotropies, all
assuming $\Omega_b = 0.1$ and $h=0.5$.
The curves peaked at $l=4000$ and $l=20000$ are from the uncorrelated
$R=5$ Mpc and $R=1$ Mpc models respectively.  The oscillations
are ringing due to the sharp cutoff of $V(r)$.  The dashed curves
show the model of Eq.~\ref{eqn:2pt_pk} for $z_i=26$ (bottom) and 
$z_i=31$ (top). The remainig solid lines include
velocity-ionization field cross-correlations.
The correlated model curves do damp beyond $\ell\simeq 1/R > 30000$.
}
\label{fig2}
\end{figure}

{\parindent0pt\it Correlated Inhomogeneous Reionization.}
Because the ionizing sources in the model of Eq.~\ref{eq:uc2} and that 
devised by GH are independent of each other,
correlations in the ionization field only exist 
on scales comparable to, or smaller than, the ionization radius $R$.
However, overdense
regions, such as galaxies and clusters of galaxies, 
are observed to be highly correlated with each other.
If overdense regions are the sources of the ionizing 
radiation (which they are very likely to be) then their clustering
will lead to long-range correlations in the ionization field
and therefore affect $C^{\rm IHR}(\theta)$.

The physical process we imagine is that the mass in any region
where the linear theory density contrast smoothed on a comoving scale $R_c$,
$\delta_{R_c}$, exceeds a critical threshold ($\delta_c=1.69$),
collapses, forms stars or quasars which then ionize a region with 
size $R=E^{1/3}R_c$. The efficiency, $E$, is the
ratio of the volume of the ionized regions to collapsed regions.
From this scenario we expect\cite{BBKS,NK}
\bea
\bar x_e &=& 2 E \int_{\delta_c}^\infty d\delta P_{R_c}(\delta)= 
E\ {\rm erfc}\Big(\frac{\d_c}{\sqrt{2}\ \sigma_{R_c}(\eta)}\Big), 
\label{eqn:1pt_pk}\\
C_{x_e}^{\rm co} &=&\langle x_e x_e^\prime \rangle = 
\alpha \alpha'\int_{\delta_c}^\infty d\delta
\int_{\delta_c}^\infty d\delta' \  P_{R_c}(\delta,\delta') \nonumber \\
&\approx & \frac{\alpha \alpha'}{2\pi}
\frac{\Xi_{R_c}(r)}{\sigma_{R_c}(\eta)\sigma_{R_c}(\eta')} 
\exp\Big[-\frac{\nu^2(\eta)+\nu^2(\eta')}{2}\Big],
\label{eqn:2pt_pk} 
\eea
where ${\rm erfc}(x)$ denotes the complementary error function, 
$\alpha \equiv 2 E(1-\bar x_e)$ (see below), $P_{R_c}(\d)$ and 
$P_{R_c}(\delta,\delta')$ are the one and two-point distributions
of the smoothed field $\d_{R_c}$, respectively; the smoothed 
correlation function is given by $\langle \delta\delta'\rangle 
\equiv \Xi_{R_c}(r)$, $\sigma_{R_c} \equiv \Xi_{R_c}(0)$, and $\nu(\eta)
\equiv \d_c/\sigma_{R_c}(\eta)$.    

Despite a number of studies of the 
details of reionization \cite{CDMIR,TSB,HL}, we are 
still far from a satisfactory understanding. Therefore, we 
take two simple models for $E$, which can 
vary by many orders of magnitude (see e.g. table 2 in\cite{TSB}).  
The requirement that objects more 
massive than $M_{c}(z) \simeq 10^{8}\msun [10/(1+z)]^{3/2}$ are 
necessary for reionization leads to a smoothing scale of 
$R_{c}=0.22 \eta$ Mpc\cite{HL}.  Our two choices for the efficiency
are $E=7\times 10^{5}\ \eta^{6}$ \cite{HL}, resulting in $z_{i}=26$ and 
$E=114$ (the ``middle of the road'' model in 
\cite{TSB}), which gives $z_{i}=31$.
Note that current 
scenarios for reionization generally lead to $E\gg1$, i.e. the sources 
tend to ionize a region much larger than their typical volume 
(see however\cite{pj98} for a different view).  


Fig.~1 shows the results of Eq.~\ref{eqn:2pt_pk} for the $z_{i}=26$
model. The clustering of overdense regions naturally leads to a much
wider $C_{x_e}^{\rm co}$, which falls off only as a power-law, thus
more efficiently modulating $C_v$. For this reason, the correlated
models in Fig.~2 (dashed lines) show a much wider distribution of
power, and the white noise regime is only reached at scales
much larger than the patch size.  
Over the range of $\ell$
plotted, the power spectrum of the uncorrelated models with the same
values of $R$ as in the peaks model would be orders of magnitude
smaller. 
The dashed lines in Fig. 2 are the result of assuming 
$\langle x_e x_e^\prime v v'\rangle=
\langle x_e x_e^\prime\rangle\langle vv'\rangle$ and
then applying Eq. 13 for $\langle x_e x_e^\prime\rangle$.  
For the solid lines, 
we use an expression for $\langle x_e x_e^\prime vv'\rangle$ 
similar to that for 
$\langle x_e x_e^\prime \rangle$ in Eq. 13 but
with an integral over the joint probability 
distribution of $\delta,\delta',v,v'$\cite{KSD2}. 
This leads to a modification of the white
noise behavior at large scales.  
Cross-correlation between HR and IHR 
leads to higher-order terms such as 
$\bar x_e \langle x_e^\prime\d vv'\rangle$.  
These are of comparable magnitude to 
$\langle x_e x_e^\prime vv'\rangle$ and 
will be discussed in \cite{KSD2}.  Note that the flatness
of the power spectra in Fig. 2 above $\ell \simeq 3000$ 
reflects the near scale-invariance
of the matter power spectrum over the relevant length scales.

{\parindent0pt\it Impact on Parameter Determination.}
The anisotropies produced during IHR
may significantly affect the exquisite parameter determination 
anticipated from future CMB observations. Suppose a
multi-dimensional fit is performed on the data to extract
a set of parameters $\{p_i\}$. If the fit assumes
$C_l(\{p_i\})$ that ignores the 
contribution from $C_l^{\rm IHR}$, then
each parameter will be incorrectly estimated by an amount
\be
\Delta p_i = F^{-1}_{ij} \sum_l w_l\ 
{\partial C_l\over \partial p_j}\ C_l^{\rm IHR} 
\ee
where the weights $w_l$ are the inverse of the squares of
the errors expected on $C_l$'s (the errors are the sum
of the errors due to sample variance and those due to
noise \cite{forecast,BET}); $F_{ij}$ is the Fisher matrix $\sum_l w_l
(\partial C_l/\partial p_i) (\partial C_l/\partial p_j)$. 
Table 1 shows the ratio of this systematic offset to the 
statistical uncertainty for Planck\cite{satellites}.

{\parindent0pt\it Sources of Uncertainty.}  Even within the context of
our peaks model, Eq.~\ref{eqn:2pt_pk} is not exact.  Since the ionized
region is larger than the region that collapsed (i.e., $E>1$) a more
rigorous approach would be to calculate the probability of two points
both being within a distance $R$ of one or more peaks.  Instead, we
have calculated the probability that two points are both {\it within}
a peak and then scaled the resulting two-point function by $EE'$.  We
expect this approximation to be best for $r > R$, which covers the
range of relevance for the curves in Fig. 2.
Because we have not taken the more rigorous approach, we must include
suppression factors to correct for overcounting overlapping regions.  These
are the factors of $(1-\bar x_e)$, motivated by their appearance in
our fully tractable uncorrelated model.  Removing them boosts the
power spectrum by a factor of about 5.

Uncertainty in $E$ has a much milder
effect on uncertainty in $C_l^{\rm IHR}$ than one might expect from a
quick glance at Eq.~\ref{eqn:2pt_pk}.  
The reason is that
increasing $E$ causes reionization to occur earlier and therefore
decreases the exponential factor in Eq.~\ref{eqn:2pt_pk}. 
We find that decreasing $E$ by a factor of 10 decreases
$C_l^{\rm IHR}$ by a factor of two.

Our peaks model, and the calculations which lead to estimates
of its parameters, are highly idealized.  Although more
sophisticated studies will eventually alter the details of our 
results, we believe
significant power at scales larger than the patch size is a
natural consequence of cosmic structure formation via 
gravitational
instability.

\acknowledgements 
We thank J.R. Bond, A.H. Jaffe, D. Pogosyan, A.
Stebbins and M. Zaldarriaga for useful discussions, 
and JRB for the $C_l$ derivatives.
SD is supported by the DOE and by NASA Grant NAG 5-7092.

\begin{table}
\begin{tabular}{|c|c|c|c|c|c|c|c|}
model& $n_s$ &$r_{ts}$ & $\tau$ &$\Omega_b h^2$ & $\Omega_{\rm vac}h^2$ & $\Omega_{\rm m} h^2$ &  $\Omega_\nu h^2$\\ \hline
$z_i=31$ & 0.75 & 0.65 & 0.01 & 2.25 & 0.20 & -1.02 & 0.94\\
$z_i=26$  & 0.34 & 0.29 & 0.006 & 1.01 & 0.09 & -0.46 & 0.42
\end{tabular}
\caption{Ratio of systematic offset to statistical uncertainty
expected for Planck \protect\cite{satellites} from the two ``peaks'' models 
in Fig. 2.  Since IHR affects very small scales, 
MAP \protect\cite{satellites} is much less sensitive to this offset.
Variables are, from left to right, the primordial
power spectrum index, the ratio of tensor to scalar contributions
to $C_2$, the optical depth and the contribution to $h^2$ from 
(b)aryons, (curv)ature, cold dark (m)atter and massive neutrinos.  
The power spectrum derivatives 
were evaluated at the parameter values of the SCDM model 
in \protect\cite{BET}. This does not include polarization
information; however, since polarization is sourced
by the quadrupole moment, not velocity, the polarization 
IHR power spectrum will be down by roughly a factor of $(10^{-5}/10^{-3})^2$
from the IHR temperature power spectrum.
}
\end{table}

\end{document}